# Lattice Boltzmann Computation of Plasma Jet Behaviors : part II Argon-Nitrogen Mixture

R. Djebali [1,2*], B . Pateyron [2], M. El Ganaoui [2], H. Sammouda [1]

**Abstract** – *In this paper an innovative computational approach, namely the Lattice Boltzmann Method (LBM), is used for simulating and modeling plasma jet behaviors. Plasma jets are a high temperature flows, then all physical parameters are temperature dependent. This work aims to address the issue of simulating plasma-jet from the point of view of extending the applications to simulating flows with temperature-dependent diffusion parameters (viscosity and diffusivity), focusing on the phenomena occurring in plasma-jet flow for a mixture of plasma gases, N2-Ar62.5% vol. Argon and Nitrogen are two gases of the most ones used in plasma spraying. The mixture is used when looking for some jet properties. We limit our effort to take out the dynamic and thermal characteristics of this complex flow using the lattice Boltzmann equation. An important section focuses mainly on the validation of our results with compute jet dynamics software such as GENMIX and Jets&Poudres developed in laboratory SPCTS in several updated edition. These codes established for many turbulence models (k-epsilon, k-omega, Prandtl's models…) are helpful numerical keys for understanding the physics of plasma jets and plasma spraying. Our numerical results based on the centerline temperature and velocity profiles, its distributions over the computational domain, the gaussian radial profiles and the effects of inlet quantities are analyzed. The quality of the results shows a great efficiency for the lattice Boltzmann method.*
**Keywords**: *Lattice Boltzmann Method, Modeling and simulation, Plasma jet*

## I. Introduction

Plasma jets are used mainly for the spraying, decomposition, and synthesis of new materials. The use of the plasma jet strongly extended the technological possibilities to any material that could melt. The plasma jets produced discharging at pressures close to atmospheric one are characterized by the high temperatures (around 5000-15000 K, far above the melting temperature, and vapor temperature of any known material) of heavy species and high velocities (between 100 m/s and 2500 m/s) of plasma flow. Hence, a complex flows under these conditions.

Numerous experimental and numerical efforts are conducted in this subject to reach high performances (of surfaces treatments and coating), for economic constraints and to well understand the complex heat, momentum and mass transport coupling. This is because plasma temperature and flow fields, in the flow core, affect absolutely the in-flight particles trajectories, and their temperature histories and then the quality and the formability of thermal spray.

Dealing with plasma jet, in one side, as former studies, E. Pfender et al. [1] have performed a simulation of Argon plasma and compared their results to available experimental measurements.

These studies and others ones deal with 3D, 2D, and 2D1/2(axisymmetric), with and without swirling-velocity in both laminar and turbulent regimes [2]-[6]. It has been shown, then, that thermal and dynamic behaviors of plasma jets depend on a great deal of parameters that interacts starting from burner chamber to the coating formation. In a comparative study, D.-Y. Xu et al. [7] have shown that using argon instead of air as surrounding gas of a laminar argon plasma jet avoids undesired oxidation of metallic materials and increases the length of jet high-temperature region and the mass flow rate but decreases the gas specific enthalpy in the jet downstream region. For this reason the surrounding gas we use will be the same as for the plasma jet.

In other side, industrials seek for new alloys that serve well in many fields, such as firebox parts and aero-engine. That's why many gas mixtures are investigated and numerous powder ingredients are tested and employed. The excellent choice will be the response of efficient numerical studies and the results of experimental tests. In this work we study the characteristics of a mixture of gases, the N2-Ar 62.5% vol.

It is worth noting that argon plasma jet is at the head of plasma gas nature investigated, and that most authors

**Nomenclature**

| | | | |
|---|---|---|---|
| $c_s$ | lattice sound speed | $Ma$ | Mach number |
| $f_k, g_k$ | discrete distribution functions for density and temperature | **Greek symbols** | |
| $f_k^{eq}, g_k^{eq}$ | equilibrium distribution functions parts | $\Delta$ | filtering width($=\Delta x=\Delta y$) |
| $H$ | domain height | $\delta_{ij}$ | Kronecker symbol |
| $L$ | domain length | $w_k$ | weighting factors |
| $R$ | Jet radius | $\rho$ | $\sum_k f_k$ fluid density (volumetric mass) |
| $T$ | dimensionless temperature field | $\upsilon$ | kinetic viscosity |
| $e_k$ | discrete lattice velocity | $\alpha$ | thermal diffusivity |
| $x$ | lattice node in $(x,y\equiv r)$ coordinates | $\tau_\upsilon$ | relaxation time for the velocity field |
| $u$ | $=(u_x, u_r)$ axial and radial velocity components | $\tau_\alpha$ | relaxation time for the tempearture field |
| $F$ | external forcing terms $F_1$ and $F_2$ for density | **Subscripts Suscripts** | |
| $S$ | sink term for temperature | $min$ | minimum |
| $p$ | $\rho c_s^2$ ideal gas pressure | $max$ | maximum |
| $t$ | time | $eq$ | equilibrium part |
| $\Delta t$ | time step | $i, j$ | lattice vector components |
| $\Delta x$ | lattice spacing units ($=\Delta y$) | $k$ | discrete velocity direction |
| $Pr_t$ | turbulent Prandtl number | $t$ | turbulent |
| $m$ | lattice size in radial direction | $tot$ | total quantity |
| $C$ | Smagorinsky constant | $FD$ | Finite Difference Method |
| $C_p$ | specific heat at constant pressure | $LES$ | Large Eddy Simulation |
| $Cs$ | physical sound speed | $LB$ | Lattice Boltzmann unit |
| $\overline{S}_{ij}$ | large scale strain rate tensor | $Ph$ | Physical (real) unit |

employed a two-dimensional or pseudo-three-dimensional models [8]. Often the $K-\varepsilon$ turbulent models are employed [9], however such models introduce large errors (comparatively to DNS and LES turbulence studying) that can be damped into the viscosity as it is noticed in [10]. Some others 3D studies are performed by using a commercial computational fluid dynamics package [11-12]. In other side available modeling works are almost all based on the steady flow assumption in a time-averaged sense [8], [9], [11]. However it has been shown in [13] that the plasma jet is unsteady.

In the last decades, the Lattice Boltzmann (LB) method is considered versus classical approaches to solve complex problems of heat and fluid flow [14]-[25]. Its time-dependent scheme is in accordance with unsteady plasma jet nature. In addition, the LB equation is particularly (fundamentally) adopted to simulating gas flows, which present a collisional process.

The present paper is a new approach to reach a fully LB-understanding of the underlying physical processes and characteristics in Argon-Nitrogen plasma jet. It aims also is to enrich the numerical basis in modeling the plasma dynamics.

Although of the first attempt with H. Zhang et al. [25-26] in a 2D symmetric configuration, the present work holds on a real axisymmetric configuration based on the Jian's model [27]. The model was successfully used by R. Djebali et al. [28] to simulate argon plasma jet in a LES-LBM turbulent model. The present paper is a some what extension of work of [28] to investigate mixture gases. Furthermore, it is well to mention that plasma jet is laminar in its core but turbulent in its fringes due to the high field gradients (200 K/mm and 10 m/s/mm).

## II. Numerical model

In this section, we present the problem governing equations under adequate assumptions, the lattice Boltzmann thermal axisymmetric formulation coupled with a turbulent model and the procedure to account for the temperature dependent diffusion parameters, that leads to conversion table between physical and lattice scales.

## II.1. Basic Assumptions and governing equations

The assumptions used in this study include: the plasma jet flow is time-dependent during the computation, axisymmetric and turbulent, the plasma is in the LTE and the radiation heat loss is neglected, all the plasma properties are temperature dependent, the swirling velocity component in the plasma jet can be neglected in comparison with the axial velocity, the plasma jet flow is incompressible [26], then obeys to the condition low Mach number, hence the compression work and the viscous dissipation can be neglected in the energy equation and finally the gravity effect is neglected.

Based on the above-mentioned assumptions, the continuity, momentum and energy equations in (z, r) coordinates are, in tensor form, as follows:

$$\begin{cases} \dfrac{\partial u_j}{\partial x_j} = -\dfrac{u_r}{r} \\ \dfrac{\partial u_i}{\partial t} + u_j \dfrac{\partial u_i}{\partial x_j} = -\dfrac{1}{\rho}\dfrac{\partial p}{\partial x_i} + \upsilon \dfrac{\partial^2 u_i}{\partial x_j^2} + \dfrac{\upsilon}{r}\dfrac{\partial u_i}{\partial r} - \dfrac{\upsilon u_i}{r^2}\delta_{ir} \\ \dfrac{\partial \theta}{\partial t} + u_j \dfrac{\partial \theta}{\partial x_j} = \alpha \dfrac{\partial^2 \theta}{\partial x_j^2} + \dfrac{\alpha}{r}\dfrac{\partial \theta}{\partial r} - \dfrac{\dot{w}}{\rho C_p} \end{cases} \quad (1)$$

Were $t$ the time, $u_r$ and $u_z$ are the radial and axial velocities respectively, $\rho$ is the density, $\theta$ is the gas temperature, $\upsilon$ is the kinetic viscosity, $\alpha$ is the thermal diffusivity, $C_p$ is specific heat at constant pressure, $p$ is the pressure, $\dot{w}$ is the radiation power per unit volume of plasma (ie in W/m$^3$) and $\delta_{ir}$ is the Kronecker delta function defined as:

$$\delta_{ir} = \begin{cases} 0 & \text{if } i \neq r \\ 1 & \text{if } i = r \end{cases} \quad (2)$$

## II.2. Axisymmetric formulation of the lattice Boltzmann method for incompressible fluid flows

In this study, we will use the passive scalar approach for computing the temperature field. In other side, it is well known that the most employed 2D lattice Boltzmann model is the D2Q9 one, used in square lattice. We have found that the D2Q9-D2Q4 is a suitable model for simulating thermal flows. First it is more stable then the D2Q9-D2Q9 model. Second, it preserves the computation effort.

The standard lattice Boltzmann method is used is Cartesian coordinates. The first intend to represent axisymetric flow was with Y. Peng et al. [29]. However, the temperature field was solved by using Finite Difference method. Recently, some new formulations are available [16, 22, 24, 25-29]. The Jian's model [28] will be used in this work for simplicity.

The proposed LB model can be written, for the nine velocity directions 0≤k≤8, as follows:

$$f_k(\mathbf{x}+\Delta\mathbf{x}, t+\Delta t) - f_k(\mathbf{x}, t) = -\dfrac{1}{\tau_\upsilon}[f_k(\mathbf{x}, t) - f_k^{eq}(\mathbf{x}, t)] + \Delta t\, F_1 + \dfrac{\Delta t}{6} e_{ki} F_{2i} \quad (3)$$

The equilibrium part $f_k^{eq}$ of distribution function $f_k$ is $f_k^{eq}(\mathbf{x},t) = \omega_k \rho [1 + 3\,\mathbf{e_k}.\mathbf{u} + 4.5(\mathbf{e_k}.\mathbf{u})^2 - 1.5 u^2]$ and $\tau_\upsilon$ is linked to the kinetic viscosity as $\upsilon = \dfrac{\tau_\upsilon - 0.5}{3}\dfrac{\Delta x^2}{\Delta t}$, Further reading on the model can be found in [28].

For heat transport, the temperature evolution equation in the four-speed (D2Q4) lattice Boltzmann model is given, for 0≤k≤4, by [30] as:

$$g_k(\mathbf{x}+\Delta\mathbf{x},t+\Delta t) - g_k(\mathbf{x},t) =$$
$$-\frac{1}{\tau_\alpha}[g_k(\mathbf{x},t) - g_k^{eq}(\mathbf{x},t)]$$
$$+0.25\Delta t \frac{\tau_\alpha - 0.5}{\tau_\alpha} S \quad (4)$$

Where $S = \frac{\alpha}{r}\frac{\partial T}{\partial r} - \frac{\dot{w}}{\rho C_p}$ and can be solved by simple FD scheme, and $\tau_\alpha$ is defined as: $\alpha = \frac{\tau_\alpha - 0.5}{2}\frac{\Delta x^2}{\Delta t}$.

The macroscopic variables, density and velocity, can be computed as follows:

$$\rho(\mathbf{x},t) = \sum_k f_k \quad (5)$$

$$\mathbf{u}(\mathbf{x},t) = \frac{1}{\rho}\sum_k \mathbf{e}_k f_k \quad (6)$$

$$T = \sum_k g_k + \frac{\Delta t}{2}S \quad (7)$$

For simplicity we will adopt in what follows the transformation $(x,r) \to (x,y)$, no changes will be introduced by the transformation.

For incorporating turbulence model in the lattice Boltzmann method, we adopt the common approach due to Smagorinsky [30] in which the anisotropic part of the Reynolds stress term (see [25] for more explanation on filtering operation and filtered equations) is modeled as:

$$\vartheta_{ij} - \frac{\vartheta_{kk}}{3}\delta_{ij} = -2\upsilon_t \bar{S}_{ij} = -2C\Delta^2 |\bar{S}_{ij}|\bar{S}_{ij} \quad (8)$$

In which the isotropic part $\frac{\vartheta_{kk}}{3}\delta_{ij}$ of the Reynolds stress term is indistinguishable from the pressure term, and $|\bar{S}_{ij}| = \sqrt{2\bar{S}_{ij}\bar{S}_{ij}}$ is the magnitude of the large scale strain rate tensor $\bar{S}_{ij} = \frac{1}{2}\left(\frac{\partial u_i}{\partial x_j} + \frac{\partial u_j}{\partial x_i}\right)$;

In the LBM-LES modeling, the idea is to locally adjust the viscosity by adding the eddy viscosity to the molecular one. The total viscosity obeys the following equation (for D2Q9 model):

$$\upsilon_{tot} = \frac{\tau_{\upsilon-tot} - 0.5}{3} = \upsilon + \upsilon_t = \upsilon + C\Delta^2 |\bar{S}_{ij}| \quad (9)$$

Some algebras yields to a second order equation, the solution gives

$$\tau_{\upsilon-tot} = \tau_{\upsilon-tot}(\mathbf{x},t) = \left(\tau_\upsilon + \sqrt{\tau_\upsilon^2 + \frac{18 C\Delta^2 |Q_{ij}|}{\rho(\mathbf{x},t)}}\right)/2 \quad (10)$$

where $Q_{ij} = \sum_k e_{ki}e_{kj}\left(f_k - f_k^{eq}\right)$.

Similarly for the thermal field, the relaxation time is readjusted using the new thermal diffusivity as

$$\alpha_{tot} = \frac{\tau_{\alpha-tot} - 0.5}{2} = \alpha + \alpha_t = \alpha + \frac{\upsilon_t}{Pr_t} \quad (11)$$

Where $Pr_t$ is the so called turbulent Prandtl number, usually taken between 0.3 and 1.

*II.3. Accounting the temperature dependent parameters*

As mentioned above, Argon-Nitrogen plasma jet is a high temperature flow. So that, all the physical quantities (viscosity, diffusivity, specific heat, density, sound speed, power radiation…) are temperature-dependent. The discrete data of these quantities are coded in T&TWinner by [31]-[38].To well take into account this behavior, we have to describe the way giving the transformation of the real (physical and *Ph*- indexed) quantities to its LB values (LB indexed).

In our study, the LB viscosity (and the physical diffusivity) is fitted to polynomial curves, compromising the stability condition $\upsilon_{LB} > 2.5\,10^{-3}$, so that we have to act on the quantity $\frac{m}{L_0}$. For general cases, one obtains the same dimensionless value when making adimensional a quantity $\phi$ in *LB*-space and *Ph*-space as:

$$\frac{\phi_{LB}}{LB\_scale} = \frac{\phi_{Ph}}{Ph\_scale} \quad (12)$$

Then,

$$\phi_{LB} = \frac{LB\_scale}{Ph\_scale}\phi_{Ph} \quad (13)$$

The Table 1 summarizes the conversion rules between some LB quantities and their corresponding physical values.

## III. Model and configuration

A half plan is considered as a computational domain for the axisymmetric plasma jet. The graph is mapped in Figure 1. Where H=12*R=48 mm, L=120 mm. AB is the anode thickness, then, no-slip boundary (**u**=0) condition and a fix temperature ($T_{min}$=700K) are retained. BC is a fixe temperature ($T_{min}$=700K) and free bound for the velocity ($\partial$**u**/$\partial$**n**=0) are adopted. CD is a boundary that we will describe later. OD is an axisymmetric boundary (see [24] for further details). OA is governed by the inlet condition of Eq. (14).

$$\begin{cases} u_{in} = u_{\max}\left[1 - \left(\frac{y}{R}\right)^2\right] \\ T_{in} = (T_{\max} - T_{\min})\left[1 - \left(\frac{y}{R}\right)^3\right] + T_{\min} \end{cases} \quad (14)$$

Where $u_{\max}$ and $T_{\max}$ are the velocity and temperature of the plasma jet at the torch axis, $T_{min}$, the temperature of the anode, set to *700 K*, and *R=4mm* be the jet-radius at the torch exit.

The domain sizes are as follows: $0 \leq x \leq 120\,mm$, $0 \leq y \leq 48\,mm$. The domain is mapped by a uniform computational mesh.

## IV. Results and discussion

Mostly, numerical plasma jet simulations omit work-piece. However, work-piece constitutes a different boundary condition when spraying in spite of the most taken, free boundary. In our study we consider the two cases, with and without work-piece. When taking account of work-piece, the plasma jet shows an appreciable deformation in temperature and velocity field traces when impinging upon the substrate, likes it is shown in [2] and [39].

TABLE I
Conversion between LB quantities and their corresponding physical (real) values

| Denomination | LBM context | Physical context |
|---|---|---|
| Kinetic viscosity | $\upsilon_{Ph} = (\tau_\upsilon - 0.5)c_s^2$ | $\upsilon_{Ph} = \upsilon_{LB} \dfrac{C_S}{c_s} \dfrac{L_0}{m}$ |
| Thermal diffusivity | $\alpha_{Ph} = (\tau_\alpha - 0.5)c_s^2$ | $\alpha_{Ph} = \alpha_{LB} \dfrac{C_S}{c_s} \dfrac{L_0}{m}$ |
| Velocity | $\boldsymbol{u}_{LB} = \dfrac{1}{\rho}\sum_k \boldsymbol{e}_k f_k$ | $\boldsymbol{u}_{Ph} = \boldsymbol{u}_{LB} \dfrac{\upsilon_{Ph}}{\upsilon_{LB}} \dfrac{m}{L_0}$ |
| Temperature | $\theta = \sum_k g_k$ | $T = (T_{\max} - T_{\min})\theta + T_{\min}$ |

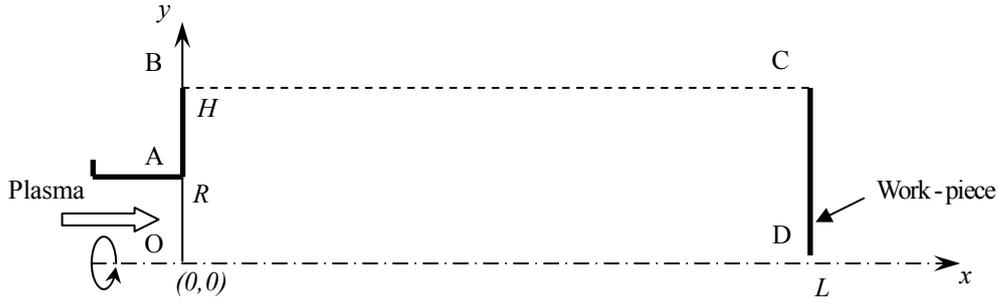

Fig.1. Computational domain

### IV.1. Validation analysis for free jet

In this case the CD edge is a free boundary and the classic extrapolation condition is adopted. The computing inlet conditions are $T_{max}$=10000 K and the velocity takes three values: 400 m/s, which serve for validation with *GENMIX* and *Jets&Poudres* software models (based on the mixing length turbulence model) [38] and 500 and 600m/s to put on view the inlet velocity effects on plasma jet behavior.

To show the ability of our thermal model to simulation axisymmetric flows, we consider, in figures 2 and 3, the present centerlines velocity and temperature distributions compared to the numerical results of the underlined softwares (with conditions $u_{\max} = 400 \,\text{m/s}$, $T_{\max} = 10000 \,\text{K}$, gaz flow rate=26 l/min corresponding to a massic flow rate of 5.2 10$^{-5}$ Kg/s, spray distance=120 mm, electric power =10 KW and efficiency =50%, gaz: N2-Ar 62.5% vol. ensuing into the same gas). The velocity and temperature profiles of our simulation compare well to the numerical results of [38]. It is well noticed that the axial temperature gradient near the inlet (interval 0-20 mm) is close to 140 K/mm (counter 195 K/mm and 167 K/mm for *GENMIX* and *Jets&Poudres* results respectively) and the velocity gradient is close to 8.4 (m/s)/mm (counter 9(m/s)/mm and 4.6(m/s)/mm for *GENMIX* and *Jets&Poudres* results respectively) which agree well with former experimental and numerical observations as noted here-above.

One can also remark that our results go well with *the* other ones. The outlying in the potential core of the plasma jet (hot zone) is probably due to the fact that ramps are used in *GENMIX* and *Jets&Poudres* codes for the inlet temperature and velocity profiles instead of ours parabolic ones. After that, in the plasma jet core, the profiles become gaussian and all the curves go together.

In the other side, it is clear that the present velocity profile occupies in the majority a mean position among the available result profiles, however I seems to be more dissipative than *Jets&Poudres* results and points well on the *GENMIX* results. That's because in *Jets&Poudres* algorithm takes in account of a kinetic energy correction (energy injected into the jet). Furthermore, we observe that the distributions of all the fields, comparatively with the experimental observations, are much the better for the *Jets&Poudres* results than the *GENMIX* ones. The form of the jet for the *GENMIX* are too expanded. We adopt for the following comparisons the *Jets&Poudres* results.

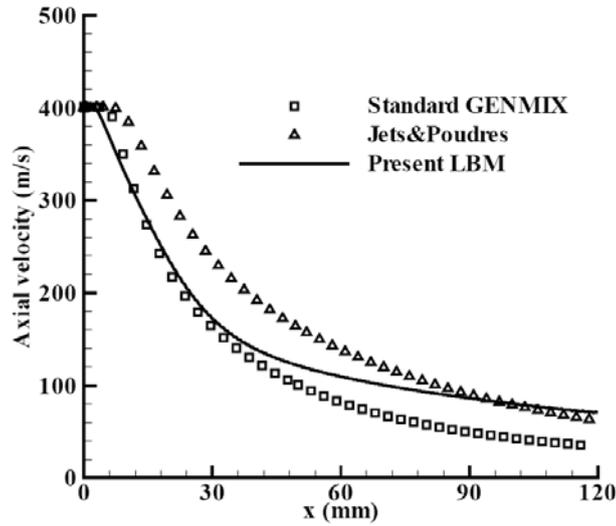

Fig.2. Centerline-temperature distribution simulated on a LBGK D2Q9 lattice with a Smagorinsky model considering $C_{smag}$=0.18 and $Pr_t$=0.3 in comparison with referenced results.

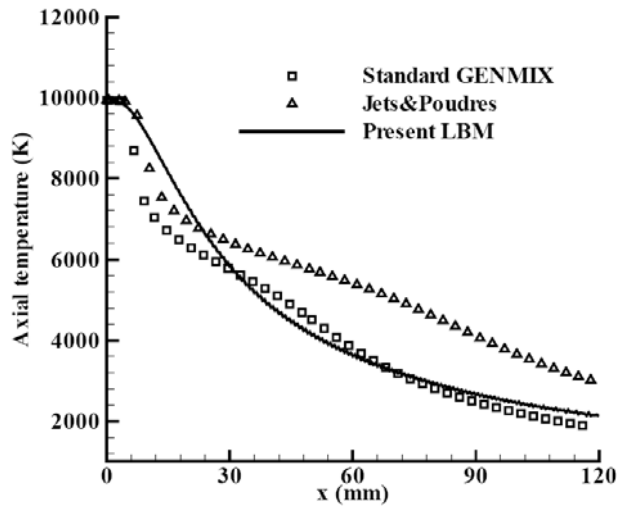

Fig. 3. Centerline-axial velocity distribution simulated on a LBGK D2Q9 lattice with a Smagorinsky model considering $C_{smag}$=0.18 and $Pr_t$=0.3 in comparison with referenced results.

Figures 4 and 5 present the isotherms and iso-axial velocities of our results and those of *Jets&Poudres*. It is clear from LB results that the temperature distribution is more expanded then the axial-velocity one, and it shares this characteristic with the Finite-Difference (*Jets&Poudres*) results. It might be mentioned that where the tempearture is higher the velocity is higher, then decreasing the probability of evaporating particles when spraying, and similarely when the tempearture is lower, the velocity is lower, then increasing the residence time of flying particles and thus continuing the melt for the solid particles core.

In simulating and modeling plasma-jets, there is no limitation in the choice of the computational domain, except the typical plasma jet length (spray distance) 100 mm, the plasma jet is observed to be fully developed for about this length. Previous studies are performed for various domain sizes in axial and radial coordinates. In this study we choose to point out the effect of enlarging the computational grille in radial coordinate on the temperature and velocity distributions. Widths 12, 24 and 48 mm are examined here for comparison on the axial distributions (the results are not presented here). The results show no big variation between 24 and 48 mm case. We will adopt the size 120*mm*x48*mm*.

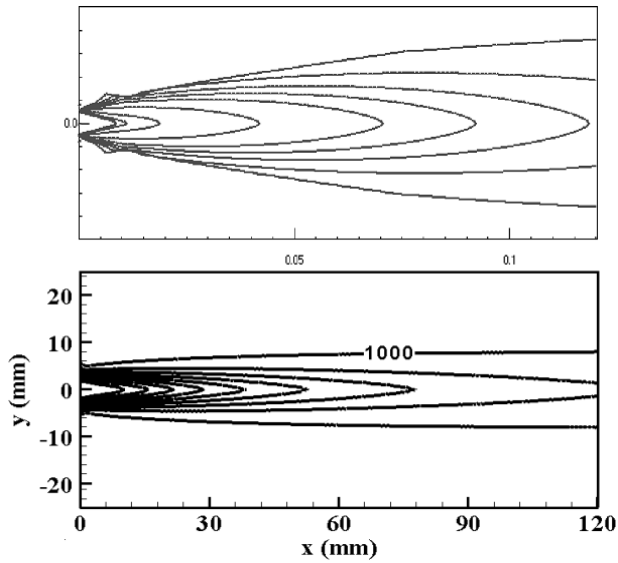

Fig.4. Isotherms traces for *Jets&Poudres* code (above) and LBGK (below) with 1000K for both the outer-line and the interval.

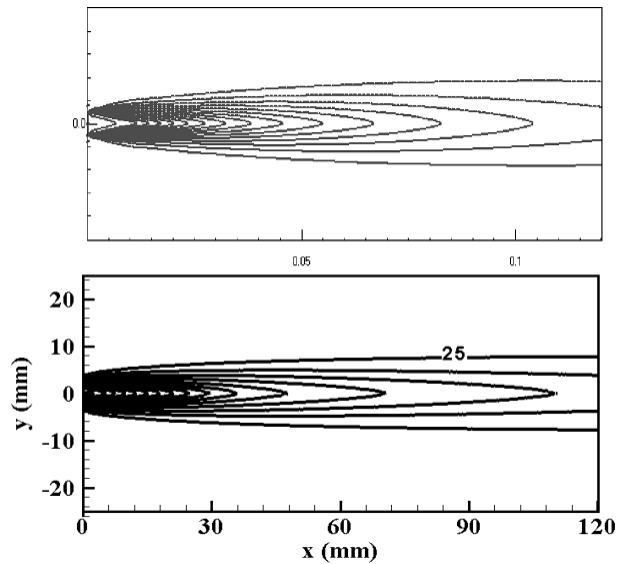

Fig.5. Axial velocity distributions for *Jets&Poudres* code (above) and LBGK (below) with interval of 40m/s and a cutoff color below 20m/s.

The velocity vectors traces of our simulation are presented in Figure 6 and are found to match the gaussian distribution radially which prove the free boundary condition taken at the north wall in spite of parabolic profiles shown in [25] which matches the non-slip boundary condition. We, also, may mention that velocity vectors traces give idea about convergence time, in our computations we found that convergence time is reached for about 50 times the number of axial grid.

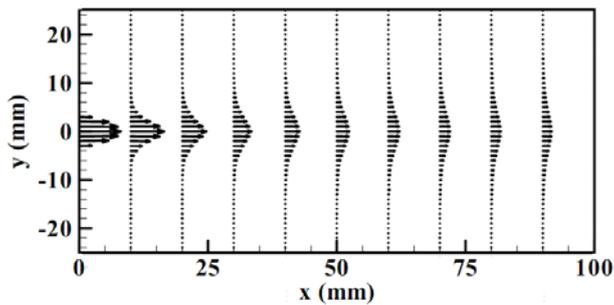

Fig.6. Velocity vectors traces for different cross sections simulated on a LBGK D2Q9-D2Q4 lattices.

Figure 7 shows the radial temperature profiles at different distances from the nozzle exit. The known "gaussian profile" is holds for all the cross sections. The maximum axial temperature decreases with increasing the axial distance, and Gaussian profile becomes more flattened.

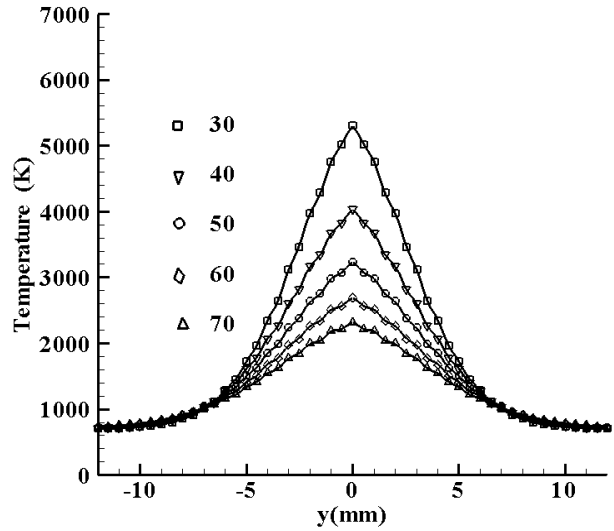

Fig.7. Radial temperature distribution for different cross sections simulated on a LBGK

To show the effects of inlet maximum velocity on the centerline velocity distribution, we perform two computations for values are 500 m/s, 600 m/s. We have to mention here that plasma jet is, however, incompressible for a Mach number is close to 0.3. For the three inlet velocities the Mach number is 0.19, 0.24 and 0.29 respectively, leading to errors close to 3.6%, 6% and 8.4% respectively (the accuracy in LB simulations are in order of $O(Ma^2)$). Figure 8 demonstrates that for high inlet velocity, the flow is entertained to the downstream region compared to *Jets&Poudres* results where no effects of inlet velocity on the downstream region. One can also say that the axial temperature and velocity gradients near high temperature keep the same above mentioned property when increasing the inlet velocity.

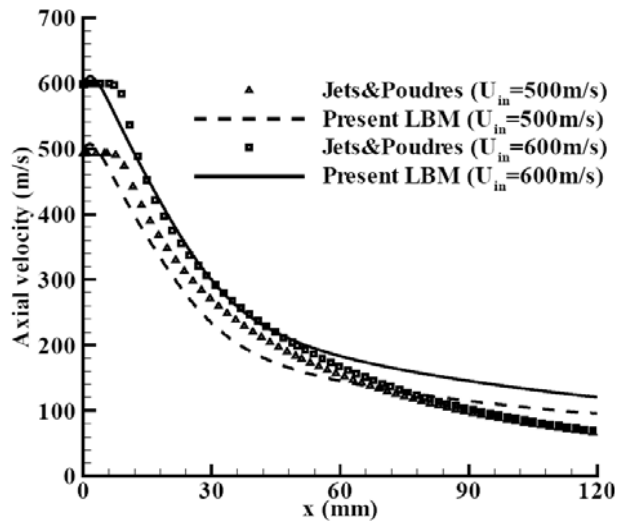

Fig.8. Effect of inlet maximum velocity on centerline-velocity.

### IV.2. Case with target (substrate)

When spraying, the target, or the substrate, constitutes a new boundary condition for the plasma jet, that is a fix wall boundary in general case. Then it is more intuitive to take in account the derived effects. This behavior have been studied later in [2], [39] and a categorical results have been demonstrated. The temperature and velocity distribution change strongly. The work-piece may have several inclinations with plasma jet axe. We just consider here the case of plasma jet impinging normally on the work-piece. The non-slip boundary condition and low temperature are retained in our treatment. The inlet temperature and velocity are chosen to be 10000 K and 500 m/s, the target stands 120 mm away from the torch exit. Results are depicted in Figures 9 and 10.

Distributions of figure 11 and 12 are in good agreement with the literature results [2], [39]. The temperature and the axial velocity distributions are flatten locally at the down stream near the work piece. The centerline fields profiles undergo a major variations. The deformation of the jet near work-piece will affect appreciably the sprayed particles trajectories and heating history and particularly its incidence.

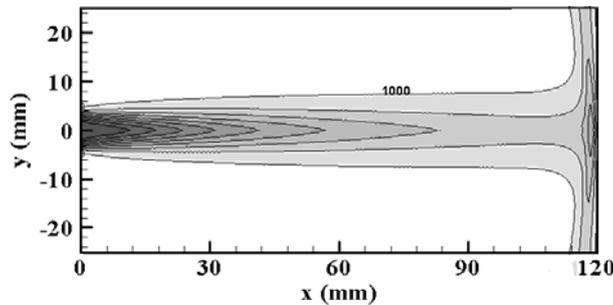

Fig.9. Temperature distribution simulated on a LBGK D2Q9 lattice with a Smagorinsky model considering $C_{smag}$=0.18 and $Pr_t$=0.3 for a jet impinging normally on the substrate with 1000K and a cutoff color below 1000K.

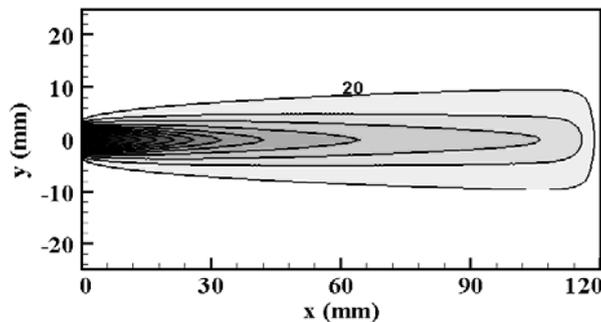

Fig.10. Axial velocity distribution simulated on a LBGK D2Q9 lattice with a Smagorinsky model considering $C_{smag}$=0.18 and $Pr_t$=0.3 for a jet impinging normally on the work-piece with 40m/s for the interval and a cutoff color below 20m/s.

## V. Conclusion

In this paper an argon-Nitrogen axisymmetric plasma-jet flowing into stagnant argon-Nitrogen is simulated by using the lattice Boltzmann method. The turbulent character is modeled and the temperature dependence of diffusion parameters is taken in account maining to important conclusions dealing with the ability of the approach to simulate complex flows; namely axisymmetric turbulent flows with strong temperature dependent physical parameters. It was shown the possible incorporation of a turbulence model.

The computed centerline temperature and axial velocity by LB method compare well with available numerical results of available software based on different turbulence models. The temperature and axial velocity distributions are more representative for the axially-extended plasma jet than other available based simulation results.

Increasing the inlet velocity leads to a translation of jet fields to downstream and increases the outlet temperature and axial velocity.

Including the work-piece as a wall boundary affects appreciably the flow structure and changes the field distributions in comparison with the free plasma jet.

# Authors' informations

1. University of Tunis El Manar, Faculty of Sciences of Tunis, Dep. of Physics, LETTM, El Manar 2092 Tunis, Tunisia, jbelii_r@hotmail.fr.

2. University of Limoges, Faculty of Sciences and Technology of Limoges, Dep. of Physics, SPCTS UMR6638 CNRS, 87100 Limoges, France.